\documentclass[manuscript]{acmart}

\usepackage{algorithmic}
\usepackage{algorithm}
\usepackage{array}
\usepackage[caption=false,font=normalsize,labelfont=sf,textfont=sf]{subfig}
\usepackage{textcomp}
\usepackage{stfloats}
\usepackage{url}
\usepackage{verbatim}
\usepackage{graphicx}

\usepackage{siunitx}
 
\usepackage{amsmath,amsfonts}

\usepackage{textcomp}
\usepackage{xcolor}
\usepackage{hyperref}
\usepackage{url}

\usepackage{xspace}
\usepackage{paralist}
\usepackage{siunitx}
\usepackage{listings}
\usepackage[para,online,flushleft]{threeparttable}
\usepackage{booktabs}
\usepackage{multirow}
\usepackage{tcolorbox}[compactbox]

\newcommand{\method}{\textrm{CANDOR}\xspace}
\newcommand{\evocandor}{\textrm{EVO-CANDOR}\xspace}
\newcommand{\methodnoplanner}{\textrm{w/o Planner}\xspace}
\newcommand{\methodnoreq}{\textrm{w/o Req.}\xspace}
\newcommand{\methodnopanel}{\textrm{w/o Panel}\xspace}
\newcommand{\evo}{\textrm{EvoSuite}\xspace}
\newcommand{\empirical}{\textrm{LLM-Empirical}\xspace}
\newcommand{\togll}{\textrm{TOGLL}\xspace}
\newcommand{\human}{{HumanEvalJava}\xspace}
\newcommand{\leetcodemedium}{{Leetcode-Medium}\xspace}
\newcommand{\leetcodehard}{{Leetcode-Hard}\xspace}
\newcommand\rev[1]{\textcolor{black}{#1}}
\DeclareSIUnit{\pp}{\textup{pp}}
\newcommand{\methodvoting}{\textrm{w/ Voting}\xspace}

\setcopyright{cc}
\setcctype{by}
\acmJournal{TOSEM}
\acmYear{2026} \acmVolume{1} \acmNumber{1} \acmArticle{}
\acmMonth{1} \acmDOI{10.1145/3803418}

\begin{document}

\title{Hallucination to Consensus: Multi-Agent LLMs for End-to-End  JUnit Test Generation}

\author{Qinghua Xu}
\email{qinghua.xu@ul.ie}
\orcid{0000-0001-8104-1645}
\affiliation{%
  \institution{Research Ireland Lero Centre for Software and University of Limerick}
  \city{Limerick}
  \country{Ireland}
}
\author{Guancheng Wang}
\email{guancheng.wang@ul.ie}
\orcid{0000-0002-4338-8813}
\affiliation{%
  \institution{Research Ireland Lero Centre for Software and University of Limerick}
  \city{Limerick}
  \country{Ireland}
}
\author{Lionel Briand}
\email{lionel.briand@ul.ie}
\orcid{0000-0002-1393-1010}
\affiliation{%
  \institution{Research Ireland Lero Centre for Software and University of Limerick}
  \city{Limerick}
  \country{Ireland}
}
\affiliation{
    \institution{University of Ottawa}
    \city{Ottawa}
    \country{Canada}
}
\author{Kui Liu}
\email{brucekuiliu@gmail.com}
\orcid{0000-0003-0145-615X}
\affiliation{
    \institution{Software Engineering Application Technology Lab, Huawei}
    \city{Hangzhou}
    \country{China}
}
\renewcommand{\shortauthors}{Xu et al.}

\begin{abstract}
Unit testing plays a critical role in ensuring software correctness. However, writing unit tests manually is labor-intensive, especially for strongly typed languages like Java, motivating the need for automated approaches. Traditional methods primarily rely on search-based or randomized algorithms to generate tests that achieve high code coverage and produce regression oracles, which are assertions derived from the program’s current behavior rather than its intended functionality. Recent advances in large language models (LLMs) have enabled oracle generation from natural language descriptions, aligning better with user requirements. However, existing LLM-based methods often require LLM fine-tuning or rely on external tools such as \evo for test prefix generation, making them costly or cumbersome to apply in practice. 

In this work, we propose \method, a novel end-to-end, prompt engineering-based LLM framework for automated unit test generation in Java. \method orchestrates multiple specialized LLM agents to collaboratively generate complete JUnit tests, including both high-quality test prefixes and accurate oracles. To mitigate the notorious hallucinations in LLMs and improve oracle correctness, we introduce a novel strategy that engages multiple reasoning LLMs in a panel discussion and generates accurate oracles based on consensus. Additionally, to reduce the verbosity of reasoning LLMs' outputs, we propose a novel dual-LLM pipeline to produce concise and structured oracle evaluations. 

Our experiments on the HumanEvalJava and more complex LeetCodeJava datasets show that \method is comparable with EvoSuite in generating tests with high code coverage and clearly superior in terms of mutation score. Moreover, our prompt engineering-based approach \method significantly outperforms the state-of-the-art, fine-tuning-based oracle generator \togll by at least 21.1 percentage points in oracle correctness on both correct and faulty source code.  Further ablation studies confirm the critical contributions of key agents in improving test prefix quality and oracle accuracy.

\end{abstract}



\received{20 February 2007}
\received[revised]{12 March 2009}
\received[accepted]{5 June 2009}

\maketitle

\section{Introduction}
\label{sec:intro}
Unit testing is a software testing activity wherein individual units of code are tested in isolation~\cite{Siddiq2024UsingStudy}. Unit testing plays a crucial role in modern software development because it helps software developers identify and fix defects in early phases~\cite{Beck2022TestExample}. However, manually creating unit tests is laborious and necessitates substantial domain expertise to craft high-quality tests~\cite{Daka2015ModelingTests, Daka2014AProblems}. Consequently, developers often forgo writing tests for their code altogether. As reported in ~\cite{Siddiq2024UsingStudy,Gonzalez2017AComprehension}, only \SI{17}{\percent} of \num{82447} studied Github projects contained test files. 

To reduce the burden of manual test writing, various solutions have been proposed to automatically generate test suites~\cite{Fraser2011EvoSuite:Software,Pacheco2007Feedback-directedGeneration, Shamshiri2015Dot,xinyi,jiahui}. A test suite typically comprises a set of test cases, where each test case is composed of a \emph{test prefix} --- inputs to the system under test (SUT) --- and a \emph{test oracle} that verifies whether the actual behavior matches the expected behavior of the SUT~\cite{Hossain2024TOGLL:LLMs}. 

Prior works in automated test generation predominantly focus on \textbf{test prefix generation}, with a goal of maximizing code coverage~\cite{Tullis2011OnLearning, Serra2019OnLater}. Traditional techniques in this area include fuzzing~\cite{Miller1990AnUtilities,Fioraldi2023DissectingEvaluation}, feedback-directed random test generation ~\cite{Csallner2004JCrasher:Java,Pacheco2007Randoop:Java,Pacheco2008FindingTesting,Selakovic2018TestLanguages,Arteca2022Nessie:Callbacks}, dynamic symbolic execution~\cite{Godefroid2005DART:Testing,Sen2005CUTE:C,Cadar2008EXE:Death,Tillmann2014TransferringDigger}, and search/evolutionary algorithm-based approaches~\cite{Fraser2011EvolutionarySuites,Fraser2011EvoSuite:Software}. Such methods are generally designed to maximize code coverage but often struggle to produce meaningful and human-readable test prefixes, primarily due to their limited understanding of the semantics of focal methods ~\cite{zhang2025large}. Therefore, recent research has shifted to leveraging large language models (LLMs) to produce more practical and meaningful test prefixes~\cite{codegeneratione}. However, ~\citet{tang2024chatgpt} and ~\citet{yang2024evaluation} found that search-based approaches like EvoSuite ~\cite{Fraser2011EvoSuite:Software} remain the best test prefix generator, substantially outperforming LLM-based approaches by approximately \SI{20}{\percent} in terms of code coverage on benchmark datasets.

In contrast to the increasing attention on prefix generation, \textbf{test oracle generation} remains underexplored. Most existing works circumvent this challenge by generating \emph{regression oracles} ~\cite{Fraser2011EvolutionarySuites,Fraser2011EvoSuite:Software,Schafer2024AnGeneration,lemieux2023codamosa,jain2025testforge,10.1145/3643769}, which are derived from the system's current behavior. \emph{Regression oracles} assume the existing code implementation is correct and simply record its behavior as the expected outcome. As such, they are primarily useful for detecting behavioral changes across software versions, not for validating functional correctness against intended specifications. 

A more rigorous and practical alternative to \emph{regression oracles} is \emph{specification-based oracles}, which are derived from specifications. However, generating correct \emph{specification-based oracles} poses significant challenges, primarily because it necessitates rigorous reasoning about program semantics and expected behaviors. To that end, several works have proposed leveraging LLMs to generate oracles from specifications. Empirical studies have shown that LLMs can achieve state-of-the-art (SOTA) performance in oracle generation for weakly typed languages like JavaScript~\cite{Schafer2024AnGeneration} and Python~\cite{wang2024chat}. However, generating test oracles for strongly typed languages such as Java remains significantly more challenging. This is due to the need for strict adherence to type constraints, language-specific syntax, and more complex execution semantics. Several works have attempted to address these challenges, using either fine-tuning (FT) or prompt engineering (PE) strategies ~\cite{chatassert,Hossain2024TOGLL:LLMs,catlm,endres2024can,Siddiq2024UsingStudy}. FT adapts LLMs to specific datasets while PE leverages off-the-shelf LLMs through carefully crafted prompts.  FT-based approaches generally yield higher oracle correctness compared to prompt-engineering-based approaches. For instance, LLM-Empirical~\cite{Siddiq2024UsingStudy}, as a representative state-of-the-art PE approach, demonstrates limited oracle correctness compared to FT-based approaches. In particular, TOGLL achieves the best oracle correctness by fine-tuning CodeParrot LLM on the SF110 dataset~\cite{Hossain2024TOGLL:LLMs}. However, we identify three key limitations of TOGLL: (1) It relies on \evo to produce initial test scaffolds, which inherits the known limitations of \evo in generating human-readable test cases and reasoning about complex programs~\cite{lemieux2023codamosa}; \citet{evosuite2014} reported side effects (e.g., ``state pollution'') on \SI{50}{\percent} of tests generated by \evo ~\cite{altmayer2025coverup}.  Moreover, \evo has not been maintained since 2021, resulting in incompatibility issues with newly-developed projects~\cite{chen2024chatunitest}. (2) It requires fine-tuning on version-specific Java data, which is costly to collect, often unavailable, and may not generalize well to new projects. Specifically, TOGLL is fine-tuned on test cases generated using EvoSuite  v1.0.6, Java 8, and Ant Builder. Applying TOGLL to other environments (e.g., EvoSuite 1.2.0, Java 11, Maven Builder) would necessitate collecting new datasets and re-fine-tuning or domain adaptation techniques; otherwise, its performance would likely degrade substantially~\cite{domain}. (3) The optimal configuration of TOGLL takes both Java docstrings and method signatures as input. \rev{Docstrings are informal natural language specifications that are ubiquitous in modern software~\cite{endres2024can}, whereas method signatures are less common in most software code bases. As demonstrated in ~\cite{Hossain2024TOGLL:LLMs}, however, the absence of method signatures leads to a \SI{20}{\percent} decrease in oracle correctness. Thus, reliance on method signatures makes TOGLL less robust when such information is incomplete or unavailable, which represents a practical limitation of the approach.}

To address the afore-mentioned challenges in test prefix and oracle generation, we propose a novel approach, \method, for automated JUnit test generation using off-the-shelf LLMs. \rev{We highlight that \method is the first multi-agent-based framework for testing Java methods using only off-the-shelf LLMs, thereby laying the foundation for future research on Java test generation.} In particular, \method aims to generate test prefixes with at least comparable code coverage and mutation scores as EvoSuite. More importantly, it aims to derive accurate specification-based oracles directly from informal Java docstrings (hereafter referred to as \emph{descriptions}), thereby compensating for the lack of formal specifications in most software systems.

For test prefix generation, \method coordinates multiple specialized LLM agents—\emph{Initializer}, \emph{Planner}, \emph{Tester}, and \emph{Inspector}—to construct and refine test files iteratively. These agents are responsible for generating an initial test scaffold, designing test plans to improve coverage, producing executable test cases, and inspecting the quality of those test cases. At this stage, oracles are generated directly from the source code, which may be faulty in practice, and thus the resulting assertions—referred to as tentative oracles—may be incorrect. 

To ensure oracle correctness, \method introduces a novel ensemble-based strategy inspired by David Hume's quote, ``\textit{Truth springs from arguments amongst friends}'', which underscores that constructive dialogue and diverse perspectives lead to robust conclusions. In this spirit, \method employs a panel discussion-style approach in which multiple \emph{Panelist} agents, powered by reasoning LLMs such as DeepSeek R1~\cite{DeepSeek-AI2025DeepSeek-R1:Learning}, independently evaluate tentative oracles against requirements derived from the natural-language description of the SUT.  Noticeably, \method requires only an informal description of the SUT's function instead of a formal specification, following the practices in prior works~\cite{Hossain2024TOGLL:LLMs,chatassert,tang2024chatgpt}. This can be attributed to recent findings suggesting that LLMs are robust to noisy natural-language instructions.  However, effective application of LLMs is non-trivial. Despite their effectiveness, LLMs often suffer from verbose and redundant outputs, known as the ``overthinking phenomenon''~\cite{Sui2025StopModels}. To address this, \method introduces a dual-LLM pipeline, where a basic LLM (i.e., without reasoning capability) agent  \emph{Interpreter}  extracts and formats key insights from each \emph{Panelist}'s output. Finally, a \emph{Curator} agent, also a basic LLM, aggregates these interpretations from the pipelines, determines necessary corrections, and generates accurate oracles through consensus. This design mitigates hallucination and uncertainty in test oracles by enabling multiple LLM agents to cross-validate tentative oracles and reach a consensus, ensuring that the final oracles align more closely with the intended behavior described in the specification.

We evaluate \method in terms of the quality of both test prefixes and oracles on two benchmarking datasets: \human~\cite{Athiwaratkun2023Multi-lingualModels} and LeetCodeJava. \human is an extensively studied dataset in the domain of automated test/code generation~\cite{jiang2023impact,Siddiq2024UsingStudy,chatassert,yuan,kang2025explainable,li2024hybrid,endres2024can,tian2025fixing}. To further assess \method's generalizability, we introduce a new dataset, LeetCodeJava, we constructed from the popular online judgement system LeetCode~\cite{2025LeetCodeDataset}, which consists of solutions to programming problems with varying difficulty levels. Experimental results show that \method is comparable with \evo in generating high-coverage test prefixes and clearly better in terms of mutation score, while producing accurate specification-based oracles. \method also outperforms the SOTA, fine-tuning-based approach \togll by at least 21.1 percentage points in terms of oracle correctness.  \rev{We remark that this comparison is inherently unfavorable to \method as it uses off-the-shelf LLMs, while \togll is fine-tuned with extra data. Fine-tuning adapts LLMs with domain-specific data, generally exhibiting higher effectiveness than prompt engineering in various tasks ~\cite{shin2023prompt,shin2025prompt,paul2023enhancing,hua2025research}. Nevertheless,  because \togll is the state-of-the-art approach for Java oracle generation, we include this comparison for completeness.}
We also assess the individual contribution of the key agents in \method. Experimental results show that removing \emph{Planner} leads to a substantial decrease in line coverage (0.050), branch coverage (0.046), and mutation score (0.070), respectively. Removing the \emph{Requirement Engineer} agent and the panel discussion decreases oracle correctness by at least 0.007 and 0.067, respectively. 

In summary, our contributions are as follows.
\begin{compactenum}
    \item To the best of our knowledge, \method is the first multi-agent LLM framework for end-to-end JUnit test generation, where specialized agents collaborate to generate, validate, and refine test cases iteratively. Unlike most existing approaches, \method does not rely on expensive LLM fine-tuning or external tools like \evo, which has not been maintained since 2021 and is difficult to adapt to new language features.
    \item We introduce a novel panel discussion-inspired strategy for test oracle generation,  mitigating LLMs' hallucination and uncertainty.
    \item To address the ``overthinking phenomenon'' of reasoning LLMs, we design a novel dual-LLM pipeline to extract concise oracle evaluations from their verbose output.   
    \item Experimental results on two benchmarks,  \human and LeetCode, demonstrate that \method compares favorably with \evo in generating high-quality test prefixes with comparable line/branch coverage and clearly higher mutation scores, while producing specification-based oracles with high accuracy, substantially outperforming a SOTA baseline by at least 21.1 percentage points.
\end{compactenum}

\rev{The rest of this paper is organized as follows. Section 2 introduces a running example that facilitates the presentation of \method in Section 3. Section 4 describes the experimental design, and Section 5 reports the corresponding results. Section 6 discusses key insights and threats to validity. Section 7 reviews related work, and Section 8 concludes the paper.}

\section{Running Example}
\label{sec:examples}

\begin{figure}[hbt!]
    \centering
    \includegraphics[width=0.8\linewidth]{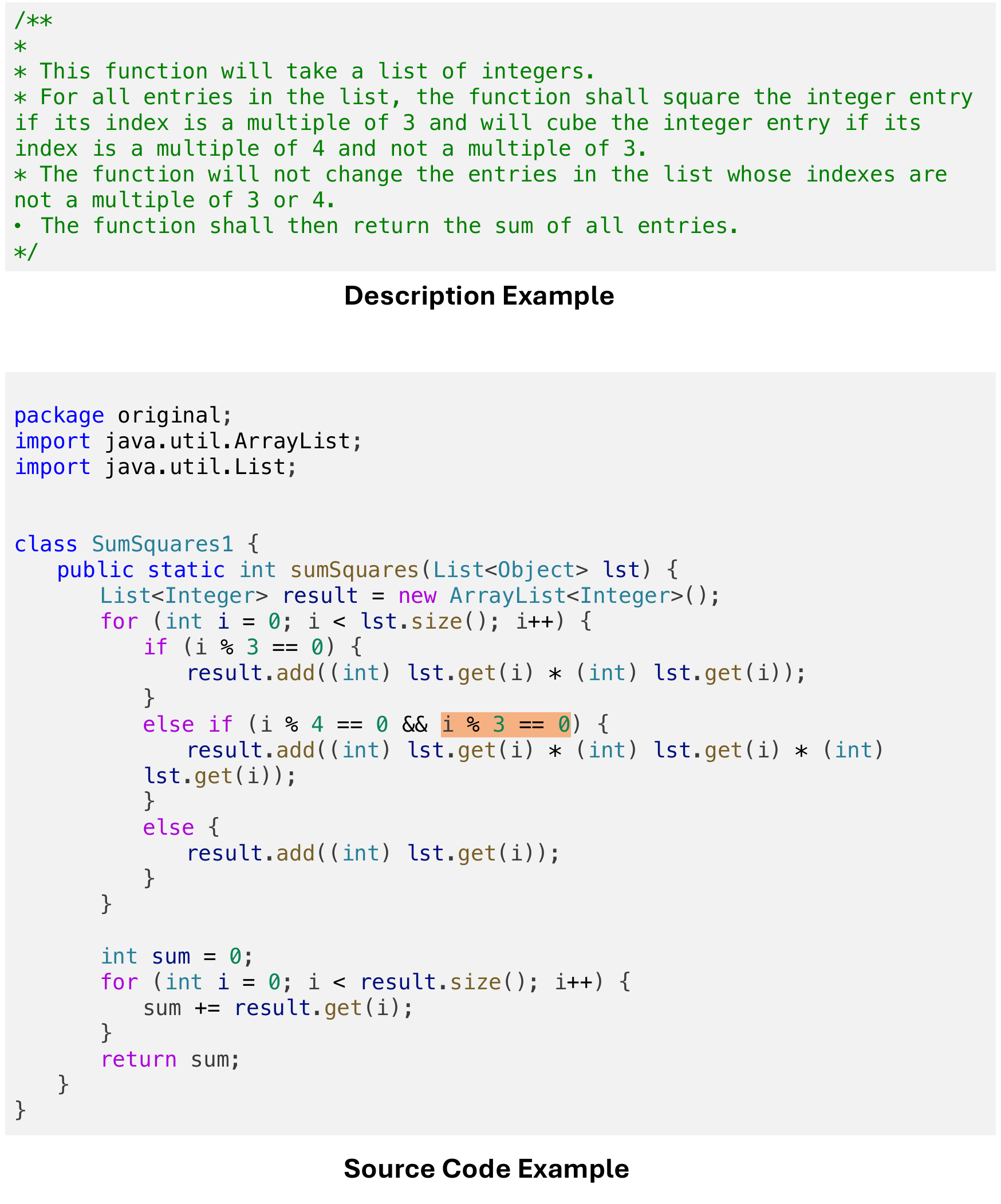}
    \caption{Running example from the HumanEvalJava dataset}
    \label{fig:example}
\end{figure}

In this section, we present the running example in Figure \ref{fig:example}, which is adapted from the HumanEvalJava dataset~\cite{Chen2021EvaluatingCode}. For each subject under test (SUT), we assume the availability of two relevant pieces of information:  the \emph{source code} and its natural language \emph{description}. 

\noindent \emph{Description.} This part provides a natural language summary of the SUT's main functionality. We assume such a description is available, as programmers typically provide such documentation explaining their code~\cite{Siddiq2024UsingStudy}. This description should include the input, the core implementation logic, and the expected output.  As depicted in  Figure \ref{fig:example}, the \emph{description}  specifies that the SUT should take a list of integers as input, implement element-wise transformation, and return the sum of the transformed list as output. For the $i$th element $lst[i]$, the transformation rule is as in Equation \ref{eq:sumsquares}.
\begin{equation}
    \label{eq:sumsquares}
    lst[i]=
        \begin{cases}
            lst[i]^2, & \text{if  } (i\%3)==0\\
            lst[i]^3, & \text{if  } (i\%3)!=0 \text{ and } (i\%4)==0 \\
            lst[i],   & \text{otherwise}
        \end{cases}
\end{equation}

\noindent \emph{Source Code.} The \emph{source code} is an implementation of the logic described in the \emph{description}, using the specified input and producing the expected output. For example, as shown in the second box of Figure \ref{fig:example}, it defines a class ``SumSquares1'' with a static method ``sumSquares''. In this example, the implementation is faulty: the condition ``$i\%3!=0$'' from Equation \ref{eq:sumsquares} is incorrectly written as ``$i\%3==0$''. 

This example represents a realistic and challenging test generation scenario, where the SUT is faulty and thus unreliable for deriving an oracle.  In real-world development, especially during early-stage or iterative coding, it is common for source code to contain latent bugs or incomplete logic~\cite{Ettles2018CommonProgrammers}. 
The goal of this paper is to generate high-quality tests with accurate oracles by combining both the \emph{source code }and its natural language \emph{description}, thus ensuring that the generated oracles are aligned with the intended functionality.

\section{Methodology}
\label{sec:method}
This section demonstrates the workflow of \method, comprising three steps: \emph{Initialization}, \emph{Test Prefix Generation}, and \emph{Oracle Fixing}. In the \emph{Initialization} step, an initial test file $v_0$ is generated from the source code and iteratively refined by fixing syntactic errors found by a validation process, resulting in a syntactically correct test file $v_1$. This file, typically containing only a small number of test cases (often $\leq 3$), serves as a template for the subsequent steps since it conforms to the syntax and conventions of Java and the testing framework JUnit 5. However, $v_1$ usually has low code coverage, as the focus at this stage is on syntactic correctness rather than test completeness. To enhance code coverage, the \emph{Test Prefix Generation} step generates additional valid test prefixes with tentative oracles, producing an expanded test file $v_2$. Notably, both Steps I and II rely solely on the source code for test generation. As a result, if the source code is faulty, the tentative oracles in $v_0$, $v_1$, and $v_2$ may also be incorrect. To address this issue, the \emph{Oracle Fixing} step conducts a panel discussion among multiple LLM agents to revise and correct the oracles in $v_2$, yielding the final test file $v_f$.

\begin{figure*}
    \centering
    \includegraphics[width=\linewidth]{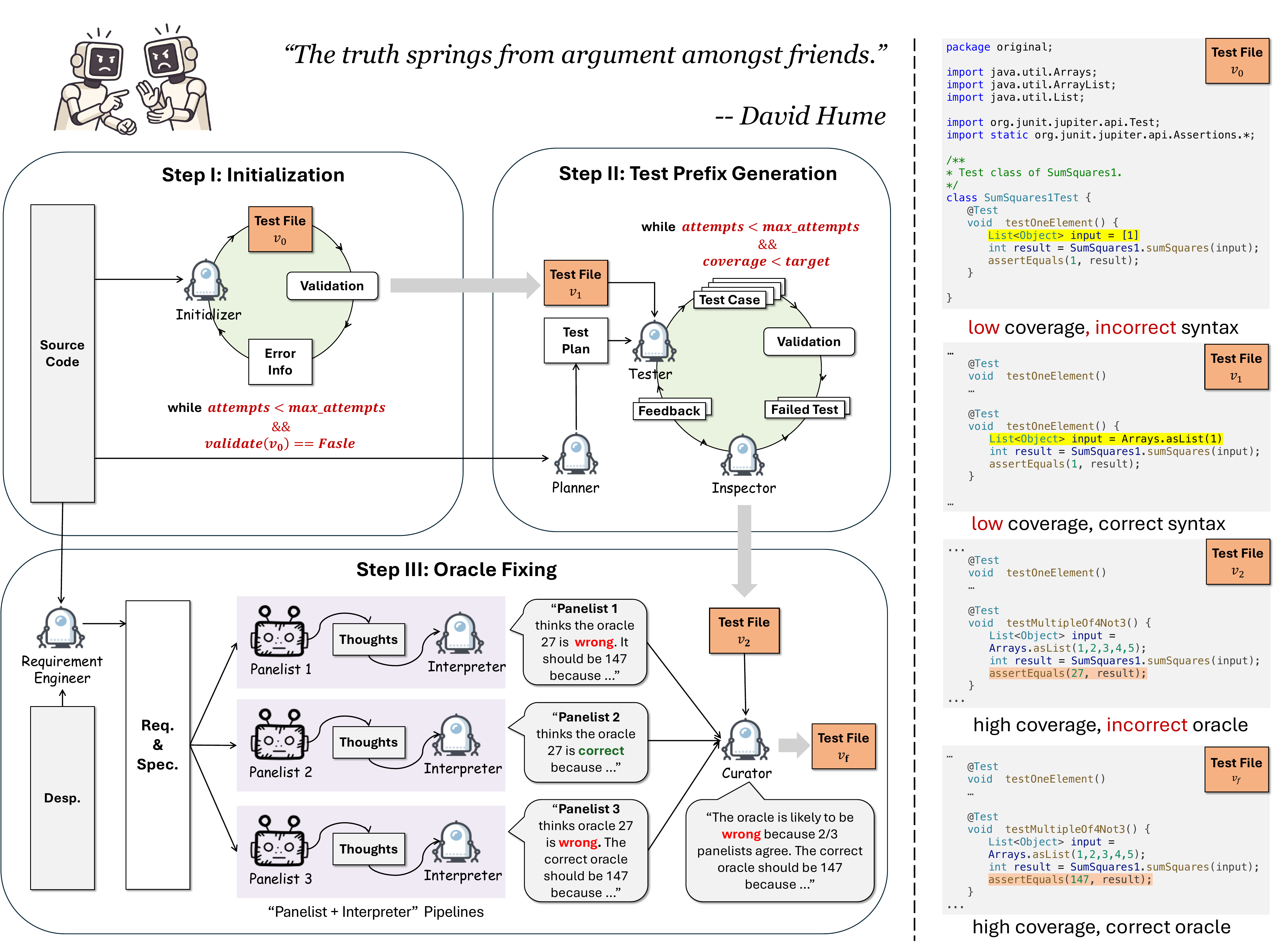}
    \caption{Overview of \method and examples. ``Desp.'', ``Req.'' and ``Spec.'' are short for "Description", "Requirement", and "Specification", respectively. }
    \label{fig:overview}
\end{figure*}
\begin{figure*}
    \centering
    \includegraphics[width=\linewidth]{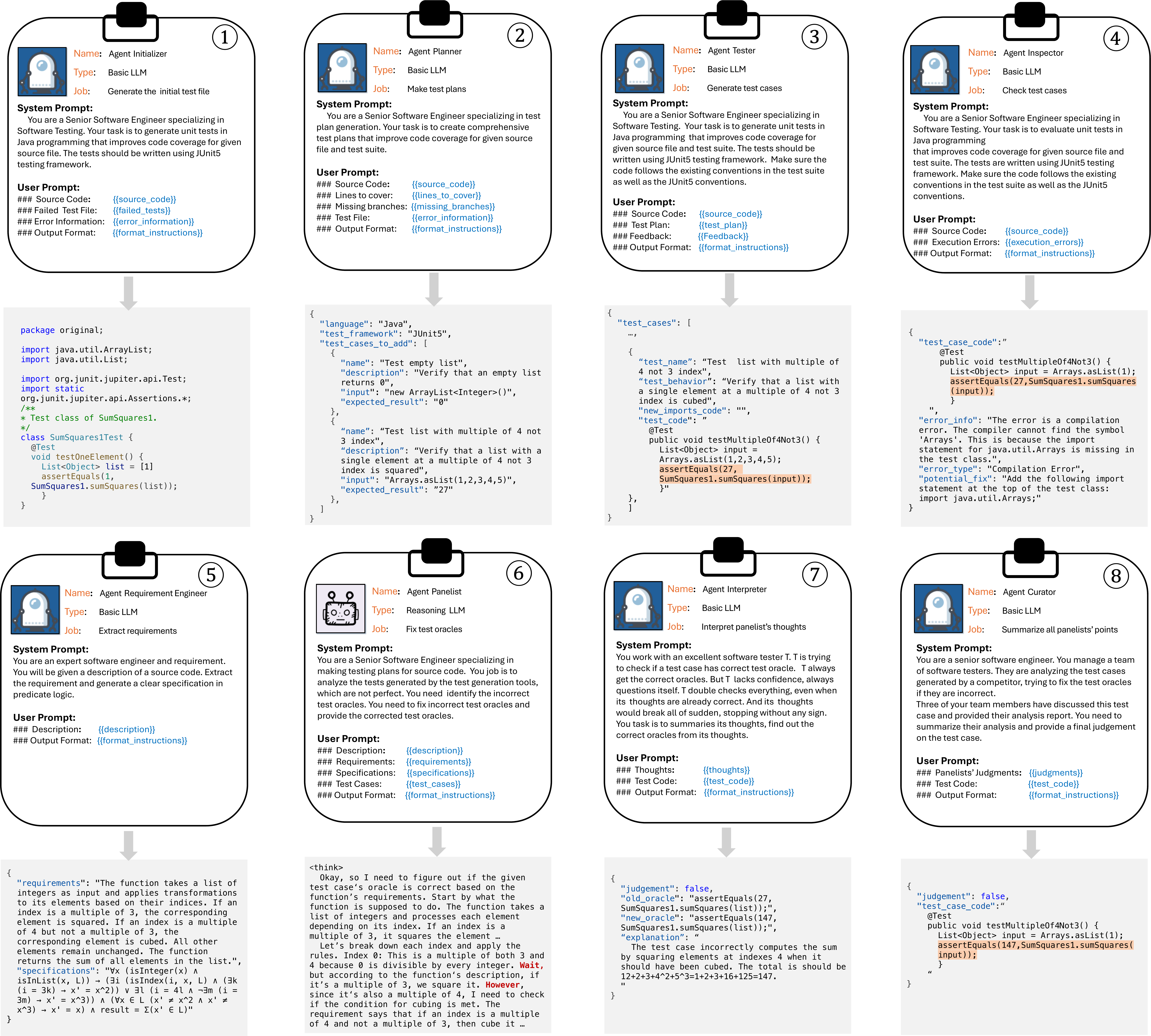}
    \caption{Profiles of LLM agents. ``Basic LLM'' and ``Reasoning LLM'' refer to LLMs without and with reasoning capability, respectively.  Each profile consists of a system prompt and a user prompt, which define the agent's role and task, respectively. Fields enclosed in double curly braces (e.g., \texttt{\{\{...\}\}}) in the user prompt represent variables that are dynamically filled for each specific case. The field \texttt{{{format\_instructions}}} is included in all user prompts and specifies the expected output format for the agent. Below each profile, an example output is provided based on the running example. \rev{The Step I, II and III of \method involve agent \textcircled{1},\textcircled{2}$\sim$\textcircled{4} and \textcircled{5}$\sim$\textcircled{9}, respectively.}}
    \label{fig:agents}
\end{figure*}
\subsection{Step I: Initialization}
\label{subsec:method-initialization}
This step aims to generate a syntactically correct initial test file. As shown in the first box of Figure 2, it involves two main components: (1) an LLM agent called the \emph{Initializer} and (2) a \emph{Validation} process. Together, they form an iterative loop to ensure the generation of a valid initial test file $v_1$. 

\paragraph{Initializer.} The \emph{Initializer} is a basic LLM agent responsible for producing an initial test file $v_0$ based on the given source code. As illustrated in profile \textcircled{1} in Figure~\ref{fig:agents}, the \emph{Initializer} is guided by two types of prompts: a system prompt, which defines the model’s role (e.g., as a software testing expert), and a user prompt, which provides specific inputs such as the source code, last failed test file, error messages, Java and JUnit 5 conventions and output format. The \emph{Initializer} responds with a JSON object containing a single field, \texttt{test\_file}, which holds the complete generated test code. Given the source file provided in Section ~\ref{sec:examples}, we present the generated initialize test file $v_0$ under profile \textcircled{1} of Figure \ref{fig:agents}, including one test case \texttt{testOneElement}. 

\paragraph{Validation.} The generated test file $v_0$ is then passed to the \emph{Validation} component, which attempts to compile and execute it. At this stage, $v_0$ often contains syntactic errors and achieves low code coverage. In the example of $v_0$ shown in Figure \ref{fig:overview}, the input list variable is incorrectly instantiated as \texttt{List<Object> input = [1]} following Python instead of Java conventions. If syntax-related errors are detected, the error messages are fed back to the \emph{Initializer} for refinement. This process is repeated until a syntactically correct and error-free test file $v_1$ is produced or a maximum number of attempts (\texttt{max\_attempts}) is reached. An example of $v_1$ is shown in Figure \ref{fig:overview},  where the input list variable is correctly instantiated as \texttt{List<Object> input = ArrayList.asList(1)} following Java conventions. While $v_1$ is syntactically valid and adheres to language and testing framework conventions, it typically still has low coverage and relies on subsequent steps to improve it. 

\rev{Note that the Initialization Step is optional in \method. If an initial test case is already provided, \method skips Step I and proceeds directly to generating test prefixes (Step II) and oracles (Step III). }
 
\subsection{Step II: Test Prefix Generation}
\label{subsec:method-prefix}
This step takes the initial test file $v_1$ from Step I and aims to enhance code coverage by generating additional test cases. As shown in the second box of Figure~\ref{fig:overview}, this step involves the \emph{Validation} process (same as in Step I) and three LLM agents: \emph{Planner}, \emph{Tester}, and \emph{Inspector}. These agents collaborate in an iterative loop to propose, generate, and refine new test cases.

\paragraph{Planner.} The \emph{Planner} is a basic LLM agent responsible for analyzing coverage gaps and proposing new test cases accordingly. As shown in profile \textcircled{2} in Figure~\ref{fig:agents}, the system prompt defines its role as a test plan designer tasked with improving code coverage. The user prompt requires inputs, including the source code, uncovered lines from the Jacoco coverage report, and the current test file. Based on this information, the Planner outputs a list of structured test plans in JSON format under the field \texttt{test\_cases\_to\_add}, with each test case described by a \texttt{name}, a brief natural language \texttt{description}, \texttt{input}, and \texttt{expected output}. A snippet of a generated plan based on the running example is provided in profile \textcircled{2},  Figure \ref{fig:agents}. This plan proposes two test cases to enhance code coverage: \texttt{Test empty list}  and  \texttt{Test list with multiple of 4 not 3 index}.   
 
\paragraph{Tester.} The \emph{Tester} is a basic LLM agent that generates executable test code based on the plans proposed by the \emph{Planner}. As shown in profile \textcircled{3} in Figure~\ref{fig:agents}, the system prompt defines the \emph{Tester}’s role as a Java unit test generator following JUnit 5 and the conventions of the given initial test file. The user prompt includes the source code, the test plan from the \emph{Planner}, and feedback from the \emph{Inspector} (if any). The output is a JSON object containing a list of \texttt{generated\_test\_cases}, with each entry specifying the \texttt{behavior}—a natural language explanation of what the test is verifying, \texttt{test\_name}, \texttt{test\_code}, and any \texttt{new\_import\_statements} required to execute this test case. A snippet of the generated tests based on the running example is provided under profile \textcircled{3} in Figure \ref{fig:agents}. The \emph{Tester} generates the test case \texttt{testMultipleOf4Not3}, with a test prefix of \texttt{ArrayList.asList(1,2,3,4,5)} and a tentative oracle \texttt{\allowbreak assertEquals\allowbreak(27, SumSquares1.\allowbreak sumSquares(input))}. Notice the tentative oracle is incorrect: it is calculated as $1+\allowbreak2+\allowbreak3+\allowbreak4^2+\allowbreak5=\allowbreak27$, whereas the correct oracle should be $1+\allowbreak2+\allowbreak3+\allowbreak4^2+\allowbreak5^3=\allowbreak147$. The fifth element was not cubed as specified in the \emph{description} due to a bug in the \emph{source code}, specifically the conditional statement \texttt{if (i\%4==0\&\&i\%3==0)} (Figure \ref{fig:example}). Additionally, the \emph{Tester} fails to identify the necessary \texttt{import} statements for \texttt{ArrayList}.   
\paragraph{Inspector.} The \emph{Inspector} is a basic LLM agent that evaluates the compiled and executed test cases to detect and explain any errors. \rev{Its role, as shown in profile \textcircled{4} in Figure~\ref{fig:agents}, is to inspect the last generated test code and its execution result and provide feedback for further improvement.} The user prompt includes the source code and error messages from the validation process. The output consists of structured feedback entries, each containing the \texttt{failed\_test\_code}, the \texttt{error\_message}, the \texttt{error\_type}, and a \texttt{potential\_fix}. This feedback is then sent back to the \emph{Tester} for refinement. A snippet of the generated feedback based on the running example is presented under profile \textcircled{4} of Figure \ref{fig:agents}.  In this case, the \emph{Inspector} correctly identifies the missing \texttt{import} statements for \texttt{ArrayList} in the test file $v_1$. However, it does not flag the incorrect oracle, as the analysis is performed against the faulty \emph{source code}, which exhibits the same buggy behavior as the test output.
 
\noindent Step II is iterative; the \emph{Planner} suggests additional test cases, the \emph{Tester} generates code, and the \emph{Inspector} provides corrective feedback. The loop terminates either when the maximum number of attempts is reached or when the code coverage satisfies a predefined threshold. This step produces the test file $v_2$, which tends to have significantly higher coverage compared to $v_0$ and $v_1$, while maintaining syntactic correctness.

\subsection{Step III: Oracle Fixing}
\label{subsec:method-oracle}
The first two steps generate test cases and tentative oracles directly from the source code. However, when the source code is faulty, the tentative oracles may also be incorrect. To address this limitation, Step III focuses on fixing the oracles using natural language \emph{descriptions}, which better reflect the intended functionality. As shown in Figure~\ref{fig:overview}, this step involves a \emph{Requirement Engineer} agent, multiple dual-LLM pipelines (\emph{Panelist + Interpreter}), and a \emph{Curator} agent. These agents collaborate to produce more accurate test oracles, even in the presence of incorrect code implementation.

\paragraph{Requirement Engineer}
The \emph{Requirement Engineer} is a basic LLM agent responsible for extracting both natural language requirements and formal specifications from the provided description of the function. As illustrated in profile \textcircled{5} in Figure~\ref{fig:agents}, the system prompt defines its role as a software engineer specializing in requirement analysis. The user prompt includes the textual description of the code, and the output consists of a set of human-readable requirements (e.g., input type and expected behavior). When possible, it also produces a formal specification expressed in predicate logic.  The grammar follows standard predicate logic and, if a formalization is not applicable or too ambiguous for the task, the specification is left empty, and only the natural language requirements are used for downstream oracle correction. The \emph{Requirement Engineer} Agent attempts to extract the specifications for every method under test and decides by itself on whether or not to provide them. The extracted requirements and specifications for the running example are shown under profile \textcircled{5} in Figure~\ref{fig:agents}.

\paragraph{Dual-LLM pipelines} Each pipeline comprises a reasoning LLM agent \emph{Panelist} and a basic LLM agent \emph{Interpreter}. Each \emph{Panelist} functions as an independent oracle corrector. As shown in profile \textcircled{6} in Figure~\ref{fig:agents}, the system prompt defines its role as a testing expert tasked with identifying and fixing incorrect test oracles generated by automated tools. The user prompt requires the natural language description, the extracted requirements and specifications from the \emph{Requirement Engineer}, and the set of test cases generated in Step II. Each agent analyzes the given test case and determines whether the existing oracle is correct. If not, it provides a corrected oracle based on the intended behavior described in the requirements and specification. To mitigate the ``overthinking phenomenon'' (see Section \ref{sec:intro}) of reasoning LLMs, each \emph{Panelist} is paired with a basic LLM agent \emph{Interpreter}. Each \emph{Interpreter} is responsible for extracting relevant insights from \emph{Panelist}'s verbose thoughts and producing structured evaluations of the tentative oracles. As shown in profile \textcircled{7} of  Figure~\ref{fig:agents}, the system prompt defines its role as an assistant working with an excellent but self-doubting tester. The user prompt requires the \emph{Panelist}'s thoughts and the test code under examination. Example outputs of both \emph{Panelist} and \emph{Interpreter}, based on the running example, are shown under profiles \textcircled{6} and \textcircled{7} of Figure \ref{fig:agents}, respectively. In this case, the \emph{Panelist} generates a lengthy, self-questioning reasoning process --- using phrases like \texttt{Wait} and \texttt{However} --- while the \emph{Interpreter} successfully extracts concise and structured oracle evaluations.

\paragraph{Curator} Multiple dual-LLM pipelines operate in parallel, each making independent judgments of the tentative oracle. The \emph{Curator} aggregates these judgments and determines the final oracle. As shown in profile \textcircled{8} of Figure \ref{fig:agents}, the system prompt defines its role as a leader of a software engineering team. While team members discuss the correctness of the tentative oracle, the \emph{Curator} is responsible for summarizing and issuing a final judgment. Its user prompt requires individual judgment from each pipeline and the test code under examination. An example output of  \emph{Curator} based on the running example is shown below profile \textcircled{8} of Figure \ref{fig:agents}. In this case, the \emph{Curator} successfully identifies the tentative oracle as incorrect and revises it into the correct oracle \texttt{assertEquals(147, SumSquares1.sumSquares(input));}.

Step III yields the final test file $v_f$, which not only achieves high code coverage—like $v_2$—but also contains corrected oracles based on the natural language \emph{description}.
\section{Experimental Setup}
\label{sec:experiment}
To assess \method, we ask the following research questions (RQs):
\begin{compactenum}[\bfseries{RQ.}1]
\item How well does \method perform at generating high coverage test prefixes? How does its performance vary under different levels of code complexity?
\item Can \method produce accurate test oracles, both when the source code is correct (i.e., conforms to the intended requirements) and when it is faulty?
\item What is the individual effectiveness contribution of key components in \method, including the \emph{Planner}, the \emph{Requirement Engineer}, and the panel discussion?
\end{compactenum}
 
RQ1 evaluates the overall effectiveness of the generated test cases by comparing code coverage and mutation scores with state-of-the-art baselines. We conduct experiments on the benchmark \human dataset~\cite{Athiwaratkun2023Multi-lingualModels}, which is extensively studied in the literature~\cite{jiang2023impact,Siddiq2024UsingStudy,yuan,kang2025explainable,li2024hybrid,endres2024can,tian2025fixing}. Additionally, we constructed a new dataset LeetCodeJava from LeetCode Platform~\cite{2025LeetCodeDataset} to assess \method's robustness across different levels of code complexity. For RQ2, we focus on the accuracy of oracle generation. We evaluate oracle correctness under two conditions: (1) when the source code is correct and conforms to the intended requirements, and (2) when the code is faulty. To simulate faulty scenarios, we use a commonly used tool (PiTest~\cite{2025PitestWebsite}) to inject mutations into the source code of both \human and LeetCodeJava to investigate \method's robustness and its practical utility in real-world testing scenarios. Specifically, we generated three mutants for each SUT. \rev{With RQ3, we assess the individual contributions of key components within \method through a series of ablation studies. Each ablation study compares \method with a variant that omits one key component, including the \emph{Planner} (w/o Planner), \emph{Requirement Engineer} (w/o Req.), and the panel discussion (w/o Panel and w/ Voting). In particular, “w/o Panel” retains only one Panelist and removes the Curator for oracle generation. “w/ Voting” replaces the panel discussion with a simple majority-voting mechanism, removing the Curator from \method. Instead of the \textit{Curator} reasoning over and summarizing the arguments of multiple \textit{Panelists}, the final oracle is directly selected based on majority agreement. We do not perform ablations on other agents, as they are essential for generating valid test cases; removing them would render \method non-functional.} \rev{To mitigate the influence of randomness, all the experiments are repeated three times.}

\subsection{Datasets}
\label{subsec:dataset} 
\rev{In this work, we focus on unit test generation for Java. But at this stage, we consider only Java methods that do not depend on user-defined classes or external libraries, thus excluding datasets such as Defects4J~\cite{defects4j} and SF110~\cite{evosuite2014}.  Addressing such dependencies requires additional techniques, such as retrieval-augmented generation (RAG) and mocking, which we plan to address in the future. As the first multi-agent-based framework for Java projects, \method lays the foundation of testing Java methods using only off-the-shelf LLMs, a significant challenge in itself. }

\rev{Specifically, we evaluate \method on two representative datasets: HumanEvalJava~\cite{Athiwaratkun2023Multi-lingualModels} and LeetCodeJava. HumanEvalJava is a representative dataset for Java unit test generation, actively studied by recent researches~\cite{jiang2023impact,Siddiq2024UsingStudy,chatassert,Chen2021EvaluatingCode,kang2025explainable,codegeneratione,li2024hybrid,tian2025fixing,endres2024can}. Further, we constructed a more challenging dataset, LeetCodeJava, to evaluate \method under scenarios with different complexity.  To demonstrate the representativeness of our selected datasets, we report key statistics in the remainder of this section, including the number of programs under test, the average line of code, and cyclomatic complexity (CC). We remark that  CC is a well-established measure of control-flow complexity, quantifying the number of linearly independent execution paths in a method~\cite{cc1,cc2,cc3,bajeh2020object}. Higher CC implies more branches to explore, thus making both test prefix and oracle generation more challenging. According to a large-scale empirical study of more than \num{862517} Java classes from \num{1000} open GitHub repositories~\cite{ccpaper}, the average CC of Java classes is \num{5.46}. }

\paragraph{HumanEvalJava} It consists of 160 Java programs implemented by Siddiq et al. to solve problems introduced in~\cite{Athiwaratkun2023Multi-lingualModels}. Each program defines a Java class under test containing a single method. The length of these programs ranges from 20 to 222 lines, with an average of 41 lines of code. \rev{The average CC of HumanEvalJava methods is 4.90, which is close to the average CC for Java methods (5.46) reported in \cite{cc}. This indicates that HumanEvalJava exhibits comparable control-flow complexity and is representative of typical Java methods. }

\rev{\paragraph{LeetCodeJava} The LeetCode Platform lists over \num{3000} programming problems, categorized by difficulty levels: ``easy'', ``medium'', and ``hard''~\cite{2025LeetCodeDataset}. More difficult problems generally require more complex code. Due to time and computational resource constraints, we randomly sampled 50 programming problems from the "medium" and "hard" difficulty categories, for a total of 100. Then, we construct the LeetCodeJava dataset by collecting their corresponding solutions~\cite{leetcoderepo}. These solutions are considered correct implementations because they are actively maintained and validated by the community. In addition to these practical considerations, we specifically chose programs of greater difficulty to ensure the dataset included more complex structural characteristics, thereby providing a more rigorous evaluation of our approach. } These methods have an average of 33 lines of code, ranging from 7 to 126. The average number of lines of code for the ``Medium'' and ``Hard'' categories is 31 and 39, respectively. \rev{The average CC of methods in LeetCodeJava is generally higher than that of HumanEvalJava  (4.90), with CC values for LeetCode-Medium and LeetCode-Hard reaching 6.30 and 9.46, respectively. These statistics highlight that LeetCodeJava-Medium includes Java classes with CC levels comparable to those observed in real-world code (\num{5.46})~\cite{ccpaper}, making it representative, while LeetCodeJava-Hard also contains more challenging cases with higher CC values. }

\subsection{Baselines}
\label{subsec:baseline}
\rev{We compare \method against four representative baselines: \evo~\cite{Fraser2011EvoSuite:Software}, \empirical~\cite{Siddiq2024UsingStudy}, TOGLL~\cite{Hossain2024TOGLL:LLMs} and \evocandor. To the best of our knowledge, \method is the first end-to-end, multi-agent-based approach for Java test generation that jointly addresses both test prefix and specification-based oracle generation. Among the baselines, \empirical is the most closely related, as it also performs end-to-end test generation but relies on a single off-the-shelf LLM. In contrast, \evo and \togll represent state-of-the-art techniques for test prefix generation and specification-based oracle generation, respectively.}

\paragraph{\evo} This widely adopted automated test-generation tool for Java programs uses search-based techniques, such as genetic algorithms, to systematically generate unit tests. It analyzes the program under test to generate test cases that maximize code coverage metrics, such as line and branch coverage. In addition to generating test prefixes, \evo automatically produces regression oracles—assertions that capture the current program behavior by checking expected outputs, exceptions, and state changes. These generated tests are structured as JUnit test cases, making \evo a popular baseline for research in automated testing and regression testing~\cite{Fraser2011EvoSuite:Software,Hossain2024TOGLL:LLMs,Tufano2021UnitContext}. As recently reported~\cite{tang2024chatgpt}, EvoSuite remains the SOTA approach in generating test prefixes in terms of code coverage, even when compared to the latest LLM-based approaches. \evo thus serves as a baseline for test prefix generation in this work.
\paragraph{\empirical} The \empirical method, proposed by Siddiq et al.~\cite{Siddiq2024UsingStudy}, leverages LLMs such as Codex, GPT-3.5-Turbo, and StarCoder to generate JUnit tests for Java programs using prompt engineering without fine-tuning. Unlike \evo, which generates regression oracles based on the current implementation behavior, \empirical generates test oracles based on natural language descriptions of the SUT. It represents the SOTA in prompt-based, LLM-driven oracle generation for Java unit tests. Among the models used in \empirical, Codex (specifically, code-davinci-002) and GPT-3.5-Turbo are the most capable. Moreover, code-davinci-002 is no longer publicly available. As a result,  we implement \empirical using GPT-3.5-Turbo. Other advanced LLMs cannot be directly applied to \empirical because their implementation—particularly the component that fixes test cases—is specifically tailored to the behavior and outputs of the LLM they use. \evo serves as a baseline for both test prefix and oracle generation in this work.
\paragraph{TOGLL} Recently, \citet{Hossain2024TOGLL:LLMs} prensented TOGLL, fine-tuning LLMs on the SF110 dataset to generate specification-based oracles. Experimental results demonstrate that TOGLL outperforms prior oracle generators by large margins in terms of oracle correctness, bug-detection effectiveness, and mutation score, setting a new SOTA in JUnit oracle generation. Additional experiments show that TOGLL achieves optimal performance by fine-tuning a CodeParrot LLM with a prompt template that includes both program descriptions and method signatures. However, since method signatures are neither commonly available in practice nor in our experimental datasets, we do not include them in the prompt. Specifically, the prompt templates include a program description, program source code, and a test prefix generated by EvoSuite. We note that comparing TOGLL and \method is biased against the latter, as TOGGL is fine-tuned on external datasets, whereas \method uses off-the-shelf, general-purpose LLMs. Nevertheless, we list TOGLL as a baseline in this work, given its SOTA performance in the oracle generation domain, while \evo serves as a baseline for test oracle generation.

\paragraph{\evocandor} \togll generates oracles based on \evo test prefixes, while \method conducts end-to-end generation, with both test prefixes and oracles generated by LLMs. Oracles for different test prefixes are not directly comparable. To ensure a fair comparison of \togll and \method in terms of oracle quality (RQ2), we consider a variant of \method, \evocandor, which generates test oracles based on the identical \evo test prefixes used by \togll.

\subsection{Evaluation Metrics and Statistical Testing}
\label{subsec:metric}
We evaluate the effectiveness of \method for test generation in terms of code coverage, mutation score, and oracle correctness. The coverage and mutation scores are calculated only for test prefixes with regression oracles to ensure a fair comparison with \evo. \rev{To reduce the influence of randomness, we repeat all the experiments three times and calculate the average of each metric. We also evaluate the significance of differences between \method and baselines by conducting statistical tests. Specifically, we compare these metrics between \method and each baseline across programs using a paired non-parametric statistical test (i.e., Wilcoxon Signed Rank test).}

\paragraph{Code coverage}This metric measures the proportion of code executed by passing tests in the generated test suites, typically reported as line coverage and branch coverage. Line coverage refers to the percentage of lines that are executed by the test suite, while branch coverage measures the percentage of control flow branches (e.g., if/else conditions) that are reached. Code coverage is a widely used metric for assessing the effectiveness of test prefixes, but it does not address test oracles~\cite{Hemmati2015HowCriteria}.

\paragraph{Mutation score}This calculates the percentage of synthetic bugs (mutants) detected by the test suite. We use PiTest with its standard mutation operators to generate mutants for each program under test. A mutant is considered killed if at least one test in the suite fails when executed against the mutated program. The mutation score is calculated as the percentage of killed mutants over the total number of generated mutants. A higher mutation score indicates a more fault-revealing test suite. 
\paragraph{Oracle correctness} This metric measures the accuracy of the assertions (oracles) in a generated test suite. Specifically, it is calculated as the number of correct oracles divided by the total number of oracles in the test suite. An oracle is considered correct if it accurately reflects the intended behavior of the program as specified by its natural language description or reference implementation. Specifically in our experiments, both HumanEvalJava and LeetCodeJava are well-established projects with no known bugs. Therefore, any failure of a generated test oracle suggests it is incorrect.

\paragraph{Statistical testing}
To compare \method against baselines, we perform the Wilcoxon Signed Rank test for all RQs, as recommended in ~\cite{Arcuri2011AEngineering}, with a significance level of 0.05. The Wilcoxon Signed Rank test is a non-parametric statistical test used to determine whether there is a significant difference between two independent distributions (e.g., method A vs. method B). 
For RQ2, we also report the $A_{12}$ effect size to measure the magnitude of difference in oracle correctness between \method and \empirical.  $A_{12}$ ranges between 0 and 1, representing the probability that method A yields better results than method B.

\subsection{Implementation}
\label{subsec:implementation}

All experiments are conducted on a Precision 7960 Tower XCTO workstation equipped with an Intel Xeon w9-3495X processor and dual NVIDIA RTX 6000 Ada GPUs. The implementation is written in Python, using the LangChain library~\cite{2025LangchainWebsite} for LLM integration. 

Most works in the literature use \evo as a baseline with its default, recommended configuration (``\texttt{assertion\_\\strategy=``mutation''; assertion\_timeout=60s}'') ~\cite{Tufano2021UnitContext,chen2024chatunitest,Hossain2024TOGLL:LLMs,Siddiq2024UsingStudy,yuan}. However, to ensure \evo is fully exercised, we experimented with different time budgets (\texttt{1}, \texttt{2}, \texttt{3}, \texttt{4}, \texttt{5}, \texttt{10}, \texttt{30}, \texttt{60}, and \texttt{120} minutes per SUT) and assertion strategies ("\texttt{MUTATION}", "\texttt{ALL}", and "\texttt{UNIT}"). We find the resulting coverage was not further improved after  ``\texttt{assertion\_timeout $>$ 2 min}'', stuck at a plateau as previously reported  ~\cite{lemieux2023codamosa}. The choice of assertion strategies did not significantly affect the quality of the generated tests. Therefore, we utilize \evo with the default assertion strategy  (\texttt{assertion\_strategy=``mutation''}) and an increased timeout threshold (\texttt{assertion\_timeout=2 min}). \rev{For PiTest, we employ the default group of mutation operators following the common practice in the literature ~\cite{shi2019mitigating,de2020using,wang2025towards}.}

\rev{We use Llama 3.1 70B as the basic LLM and DeepSeek R1 Llama-distilled 70B as the reasoning LLM. DeepSeek R1 is selected for its strong reasoning capability among open-source models. For the basic LLM, we also experimented with other commonly used alternatives such as CodeLlama 70B and Mistral 22B, however, Llama 3.1 70B achieved the best overall performance. The corresponding results are reported in the Appendix for completeness.} \rev{The max\_attempt parameter in the Initialization and Test Prefix Generation step is set to 3. If \method fails to generate an initial test file for a given method under test within these attempts, it stops further trials and proceeds to the next method.} We limit the number of output tokens of DeepSeek to 2000 for efficiency. The number of ``Panelist + Interpreter'' pipeline is set to be 3. We experimented with 1 to 5 pipelines. However, the improvements beyond 3 are not significant and incur much longer generation time. We release our code and data publicly to facilitate replication\footnote{\url{https://github.com/qiqiguana/candor}}. 
\section{Experiment Results}
\label{sec:results}
\subsection{RQ1 Results: Test Prefix Quality}
\label{subsec:rq1_results}

Table~\ref{tab:rq1_results} presents the experimental results for RQ1, comparing \method with \empirical and \evo in terms of test prefix quality, measured by line coverage (``Line''), branch coverage (``Branch''), and mutation score (``Mutation'').


\begin{table*}[hbt]
    \centering
    \caption{Experimental results of RQ1.  ``*'' denotes \method achieves significantly better results than the compared baseline.}
    \label{tab:rq1_results}
    \resizebox{\textwidth}{!}{
     
    \begin{tabular}{@{\hspace{0.5\tabcolsep}}c@{\hspace{0.5\tabcolsep}}c@{\hspace{0.5\tabcolsep}}c@{\hspace{0.5\tabcolsep}}c@{\hspace{0.5\tabcolsep}}c@{\hspace{0.5\tabcolsep}}c@{\hspace{0.5\tabcolsep}}c@{\hspace{0.5\tabcolsep}}c@{\hspace{0.5\tabcolsep}}c@{\hspace{0.5\tabcolsep}}c@{\hspace{0.5\tabcolsep}}}
    \toprule
         &\multicolumn{3}{c}{HumanEvalJava}  & \multicolumn{3}{c}{Leetcode-Medium} & \multicolumn{3}{c}{Leetcode-Hard} \\
         \cmidrule(lr){2-4}
         \cmidrule(lr){5-7}
         \cmidrule(lr){8-10}
         & Line & Branch & Mutation& Line & Branch & Mutation& Line & Branch & Mutation\\
         \midrule
        \empirical & 0.704* & 0.688* & 0.823* (2011/2443)& 0.771* & 0.732* & 0.868* (530/611)& 0.714* & 0.729* & 0.778* (668/859)\\
        \midrule
        \evo &0.961* & 0.942 & 0.858* (2096/2443)& 0.959* & \textbf{0.959} & 0.845* (516/611)& 0.984 & 0.976 & 0.888* (763/859)\\
        \midrule
        \method &\textbf{0.991} & \textbf{0.950} & \textbf{0.980} (2384/2443)& \textbf{0.990} & 0.949 & \textbf{0.939} (574/611)& \textbf{0.989} & \textbf{0.980} & \textbf{0.937} (805/859)\\
        \bottomrule
    \end{tabular}
    }
\end{table*}

\method consistently achieves high \emph{line coverage} across all datasets, with the lowest value still reaching 0.989 on the Leetcode-Hard dataset. Compared to \evo, \method shows marginal improvements, with a maximum gain of 0.031 ($0.990-0.959$ on \leetcodemedium). In contrast, \empirical performs substantially worse than \method across all three datasets, with the largest gap reaching 0.287 ($0.991-0.704$) on \human. \rev{Wilcoxon Signed Rank tests confirm that the differences between \method and \empirical are statistically significant ($p\text{-value}<1\mathrm{e}{-4}$), whereas the differences between \method and \evo are not statistically significant ($p\text{-value}>0.05$)}. 

For \emph{branch coverage}, both \method and \evo substantially outperform \empirical. The maximum branch coverage achieved by \empirical is 0.732 on \leetcodemedium, whereas the minimum branch coverage achieved by \evo or \method is 0.942 on \human. \rev{Wilcoxon Signed Rank tests confirm that the differences between \method and \empirical are statistically significant ($p\text{-value}<1\mathrm{e}{-4}$), whereas the differences between \method and \evo are not statistically significant ($p\text{-value}>0.05$).}

For \emph{mutation score}, \method performs the best across all three datasets, successfully killing 2384 (out of 2443), 574 (out of 611), and 805 (out of 859) mutants on \human, \leetcodemedium, and \leetcodehard, respectively. \rev{\method outperforms \evo by at least 0.049 ($0.937-0.888$ on \leetcodehard) and \empirical by at least 0.157 ($0.980-0.823$ on \human)}. \rev{Wilcoxon Signed Rank tests confirm that all differences between \method and the baselines are statistically significant ($p\text{-value}<1\mathrm{e}{-4}$)}. We hypothesize that the advantage of \method in mutation scores stems from the LLM's semantic understanding. Unlike \evo, which primarily evolves test prefixes to satisfy structural coverage criteria, \method also considers the semantics of the program under test (e.g., respecting input constraints, maintaining data invariants, and producing realistic usage patterns). These semantically meaningful test prefixes are more effective at revealing subtle behavioral deviations introduced by mutants, even when overall coverage is similar. Consequently, \method achieves substantially higher mutation scores than \evo. 

We also observe that \method's performance remains stable as the dataset shifts from \leetcodemedium to \leetcodehard, with only slight drops in line coverage ($-0.01$) and mutation score ($-0.02$), indicating robustness to increasing problem difficulty. We posit that this robustness originates from the pretraining of LLMs on a vast and diverse corpus, which allows \method to generalize well across programs of varying complexity in this context.   

Overall, \method outperforms both baselines across all metrics, with one exception: on the \leetcodemedium dataset, \evo achieves slightly higher branch coverage than \method by a margin of 0.01. However, the differences between \method and \evo in terms of line and branch coverage are marginal and mostly insignificant, indicating that \method achieves comparable performance in code coverage to \evo. Moreover, \method significantly outperforms \evo in mutation scores, suggesting better bug-detection capability.

\begin{tcolorbox}
Overall, \method and \evo achieve comparable, high-quality test prefixes across all datasets in terms of line coverage and branch coverage. However, \method significantly outperforms \evo in mutation scores, with improvements exceeding 0.049 on all datasets.
\end{tcolorbox}

\subsection{RQ2 Results: Test Oracle Quality}
\label{subsec:rq2_results}
\begin{figure}
    \centering
    \includegraphics[width=\columnwidth]{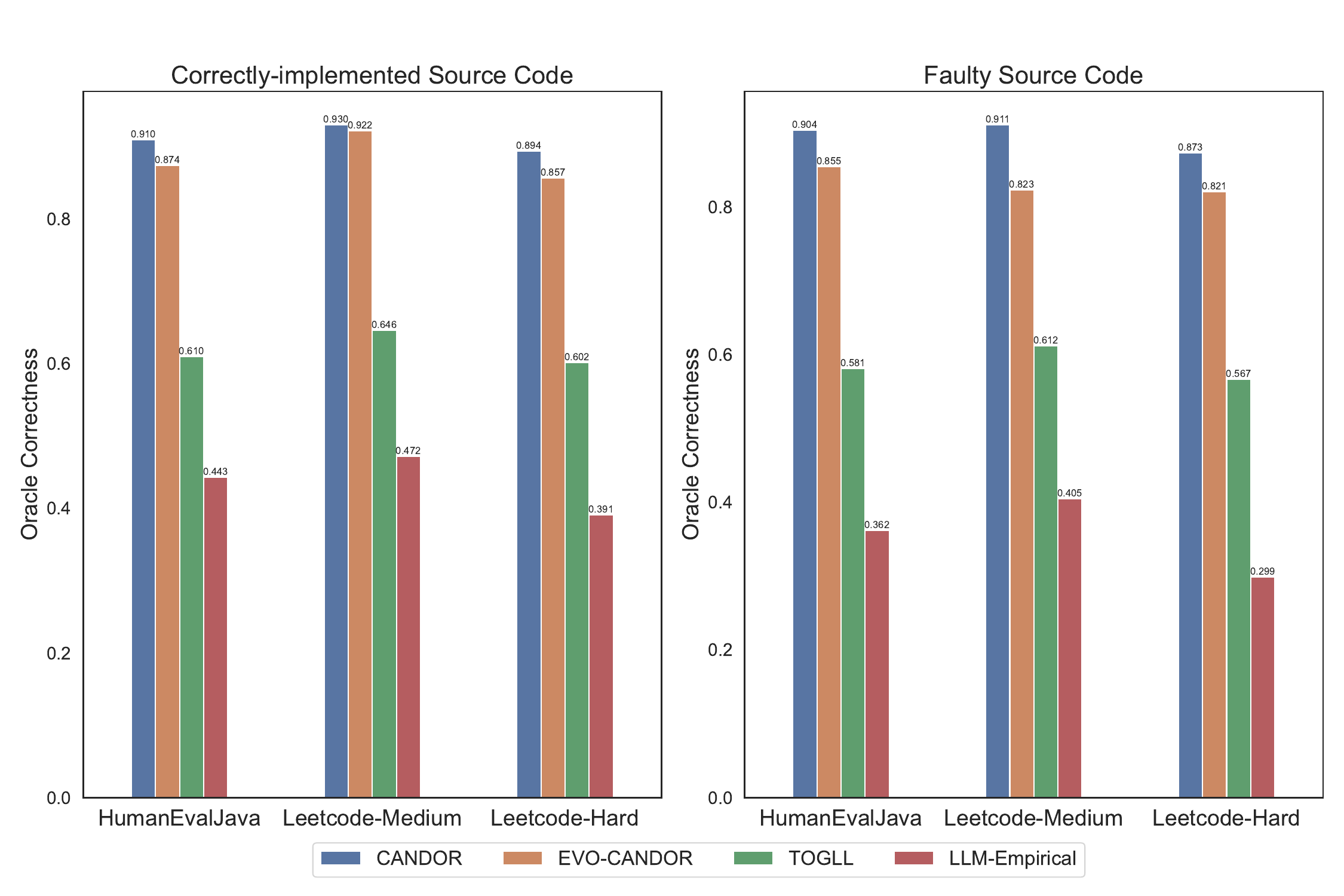}
    \caption{RQ2 results: Comparison of \method, \togll and \empirical on correctly-implemented and faulty source code in terms of oracle correctness.}
    \label{fig:rq2_results}
\end{figure}
Figure \ref{fig:rq2_results} compares the correctness of oracles generated by \method, \empirical, and \togll. We exclude \evo from this comparison since it can only produce regression oracles rather than specification-based oracles.

\noindent\textbf{On Correct Code. }As depicted in the left part of Figure~\ref {fig:rq2_results}, \method achieves the highest oracle correctness scores of 0.910, 0.930, and 0.894 on \human, \leetcodemedium, and \leetcodehard, respectively. The gap between \method and \empirical is substantial,  with a minimum of 0.467 ($0.910-0.443$) on \human in terms of oracle correctness. When compared to \togll, we refer to \evocandor's results since they both use the same \evo test prefixes. We observe that, despite the slight performance decrease compared to \method, \evocandor still outperforms \togll by a large margin, with a minimum of 0.255 ($0.857-0.602$) on \leetcodehard. \rev{We conduct the Wilcoxon Signed Rank tests for all the comparisons, and the results confirm that all differences are statistically significant ($p\text{-value}<1\mathrm{e}{-4}$).} We further calculate Vargha and Delaney’s effect size for the comparison between \evocandor and \togll and found a strong effect size ($A_{12} = 0.920$), indicating there is a \SI{92}{\percent} chance that \evocandor can outperform \togll. 

We posit that the superior performance of \method can be explained by its capability to generate oracles based on only informal descriptions. In contrast, the optimal configuration of \togll requires not only the descriptions but also method signatures, which are often unavailable in real-world projects and are absent in our datasets. Moreover, \togll is fine-tuned on the SF110 dataset, consisting of tests generated with \evo v1.0.6, Java 8, and Ant Builder. Program features beyond this environment cannot be recognized, thereby diminishing the effectiveness of \togll. For instance, we observe that \togll struggles with newer Java features, such as ``\texttt{var}'' for type inference and lambda expressions such as  ``\texttt{list.stream().map(String::toUpperCase).toList();}''   

\noindent\textbf{On Faulty Code. }As described in Section \ref{sec:experiment}, the faulty code is created by PiTest, which introduces mutations into correctly-implemented benchmarks: \human, \leetcodemedium, and \leetcodehard. As shown in the right part of Figure~\ref {fig:rq2_results}, \method remains the best approach across all datasets. Compared to \empirical, the gaps are even larger than those in correctly-implemented code in terms of oracle correctness, with a minimum of 0.506 ($0.911-0.405$) on \leetcodemedium. When comparing with \togll, \evocandor outperforms it by a large margin, with a minimum of 0.211 ($0.823-0.612$) on \leetcodemedium.  \rev{Wilcoxon Signed Rank tests confirm that all differences in the comparisons between \method and \empirical, and between \evocandor and \togll, are statistically significant ($p\text{-value}<1\mathrm{e}{-4}$)}. We also observe an even stronger Vargha and Delaney’s effect size ($A_{12} = 0.960$) in the comparison between \evocandor and \togll, as there is a \SI{96}{\percent} chance for \evocandor to outperform \togll.

The superior performance of \method on faulty code highlights its fault-detection capability. We believe such performance is achieved by deriving oracles from natural language descriptions of the source code in an effective manner, rather than relying solely on the source code itself. Specifically, the panel discussion employed by the \method enables the cross-validation on the generated oracles. This design prevents hallucination, yielding more accurate oracles than approaches that rely on single-agent reasoning.  
 
Again, it is important to recall that comparing \method with \togll is somewhat unfair, as \togll is fine-tuned with an extra dataset, while \method uses off-the-shelf LLMs. Despite this advantage, \method still outperforms \togll substantially while offering greater flexibility. Indeed, unlike \togll, \method is flexible when applied to new Java versions or system environments, eliminating the need for fine-tuning on a version-specific dataset. This is particularly important in practice as many software systems and environments are frequently upgraded, making fine-tuning costly and time-consuming.

\begin{tcolorbox}
\method generates significantly more accurate test oracles than \togll across all datasets, even though \togll is fine-tuned with additional data. Moreover, \method demonstrates robustness to faulty code by leveraging code descriptions.
\end{tcolorbox}
\subsection{RQ3 Results: Ablation study}
\label{subsec:rq3_results}
\begin{table*}[hbt]
    \centering
    \caption{Experimental results of RQ3.  ``N/A'' indicates the current ablation configuration has no influence on this metric; therefore, the corresponding experiments are not conducted. ``*'' denotes \method achieves significantly better results than the compared baseline. } 
    \label{tab:rq3_results}
    \resizebox{\textwidth}{!}{
    
    \begin{tabular}{@{\hspace{0.5\tabcolsep}}c@{\hspace{0.5\tabcolsep}}c@{\hspace{0.5\tabcolsep}}c@{\hspace{0.5\tabcolsep}}c@{\hspace{0.5\tabcolsep}}c@{\hspace{0.5\tabcolsep}}c@{\hspace{0.5\tabcolsep}}c@{\hspace{0.5\tabcolsep}}c@{\hspace{0.5\tabcolsep}}c@{\hspace{0.5\tabcolsep}}c@{\hspace{0.5\tabcolsep}}c@{\hspace{0.5\tabcolsep}}c@{\hspace{0.5\tabcolsep}}c@{\hspace{0.5\tabcolsep}}}
    \toprule
         &\multicolumn{4}{c}{HumanEvalJava}  & \multicolumn{4}{c}{Leetcode-Medium} & \multicolumn{4}{c}{Leetcode-Hard} \\
         \cmidrule(lr){2-5}
         \cmidrule(lr){6-9}
         \cmidrule(lr){10-13}
         & Line & Branch & Mutation & Oracle & Line & Branch & Mutation & Oracle & Line & Branch & Mutation & Oracle \\
         \midrule
        \method & \textbf{0.991} & \textbf{0.970} & \textbf{0.980} (2384/2443)& \textbf{0.910}& \textbf{0.990} & \textbf{0.949} & \textbf{0.939} (574/611)&\textbf{0.930}& \textbf{0.989} & \textbf{0.980} & \textbf{0.937} (805/859)& \textbf{0.894}\\
        \midrule
        \methodnoplanner & 0.892* & 0.840* & 0.869* (2125/2443)& N/A& 0.940* & 0.903* & 0.869* (531/611)&N/A& 0.935* & 0.922* & 0.833* (716/859)& N/A\\
        \midrule
        \methodnoreq & N/A & N/A & N/A& 0.882*& N/A & N/A & N/A&0.923& N/A & N/A & N/A& 0.873*\\
        \midrule
        \methodnopanel & N/A & N/A & N/A& 0.824*& N/A & N/A & N/A&0.855*& N/A & N/A & N/A& 0.827*\\
        \midrule
        \methodvoting & N/A & N/A & N/A& 0.896*& N/A & N/A & N/A&0.910*& N/A & N/A & N/A& 0.877*\\
        \bottomrule
    \end{tabular}
    }
\end{table*}

Table \ref{tab:rq3_results} compares \method with three ablation configurations: removing the \emph{Planner} agent (“w/o Planner”), removing the \emph{Requirement Engineer} agent (“w/o Req.”), and removing the panel discussion (“w/o Panel”).

Removing the \emph{Planner} significantly reduces test prefix quality across all datasets. The minimum decreases are observed on \leetcodemedium with line coverage, branch coverage, and mutation score dropping by at least 0.050 ($0.990-0.940$), 0.046 ($0.949-0.903$), and 0.070 ($0.939-0.869$), respectively. \rev{Wilcoxon Signed Rank tests confirm these decreases are statistically significant ($p\text{-value}<1\mathrm{e}{-4}$).} This underscores the critical role of the \emph{Planner}, which analyzes current coverage and strategically plans to cover missing lines. In our experiments, we observe that without the \emph{Planner}, \method generates tests exploring easy code paths in most cases, lacking a coherent strategy to explore untested code paths. 
 
Removing the \emph{Requirement Engineer} or the panel discussion primarily affects oracle correctness. In the absence of the \emph{Requirement Engineer}, oracle correctness decreases slightly by 0.028 ($0.910-0.882$), 0.007 ($0.930-0.923$), and 0.021 ($0.894-0.873$) on \human, \leetcodemedium, and \leetcodehard, respectively. \rev{Wilcoxon Signed Rank tests indicate the changes on \human and \leetcodehard are statistically significant ($p\text{-value}<1\mathrm{e}{-3}$), whereas the decline on \leetcodemedium is not significant ($p\text{-value}=0.17$).} This likely suggests that the descriptions of source code in these three datasets are clear and well-structured, which reduces the need for explicit requirement parsing using the \emph{Requirement Engineer} agent. However, in real-world scenarios with less clear documentation, this agent is expected to provide greater benefits.

\rev{In contrast, removing the panel discussion (``w/o Panel'') causes substantial and statistically significant drops in oracle correctness, i.e., 0.086 ($0.910-0.824$), 0.075 ($0.930-0.855$), and 0.067 ($0.894-0.827$) on HumanEvalJava, LeetCodeJava-Medium, and LeetCodeJava-Hard, respectively. Replacing the panel discussion with a majority voting mechanism (``w Voting'') also leads to consistent decreases across all datasets in terms of oracle correctness, with a minimum of 0.014 ($0.910-0.896$) on the HumanEvalJava dataset.  We performed Wilcoxon tests and found that all decreases are statistically significant ($pvalue<1e-2$). This aligns with our expectations, as without the panel discussion, LLMs are more prone to uncertainty and hallucination, generating plausible but incorrect oracles more frequently. In our experiments, we observed serious hallucinations during the panel discussion; in over \SI{70}{\percent} of the cases, the \emph{Panelists}' discussions contained clear disagreements. For example, a method ``\texttt{max\_element()}'' that computes the maximum of an array was mistakenly used to compute the minimum instead by one or more \emph{Panelists} (out of three). However, the \emph{Curator} can correct these errors by cross-checking the reasoning logic, inputs, and outputs of each \emph{Panelist} rather than relying solely on majority voting. This design ensures that incorrect or inconsistent outputs are identified and corrected, making the panel discussion more robust than simple majority voting in producing accurate oracles.}
\begin{tcolorbox}
Ablation results confirm the essential role of the \emph{Planner} in producing high-quality test prefixes and of the panel discussion in generating accurate test oracles. The \emph{Requirement Engineer} generally contributes to improving oracle accuracy, with its impact expected to be greater when the quality of the natural language description is low.
\end{tcolorbox}
\section{Discussion}
\label{sec:discussion}
Despite the high effectiveness of LLMs for unit test generation, instructing them to act as expected remains challenging.  In this section, we share several key insights from our experiments on effective LLM instruction (Section ~\ref{subsec:insights}), including adopting specialized agents, strategies to mitigate LLM hallucination, and methods to handle the verbosity of reasoning LLMs. We also report the limitations of our method in Section~\ref{subsec:threats}.
\subsection{Key Insights}
\label{subsec:insights}
\paragraph{Simplifying Tasks via Specialized LLM Agents} 
LLMs are increasingly studied not only due to their effectiveness but also their ease of use, as many tasks can be handled by crafting prompts for one single LLM~\cite{Guo2024LargeChallenges}. However, we found that these single-LLM approaches fall short on complex tasks, such as unit test generation. Unit test generation involves several subtasks, including planning and generating test cases. When an LLM is asked to perform both tasks simultaneously, it often becomes confused about its roles. For example, we observed that when tasked with both planning and generating Java tests, the LLM sometimes produced syntax like \texttt{[1]} instead of the correct Java syntax \texttt{Arrays.asList(1)}. We hypothesize that this happens due to interference from the planning sub-task, which is often expressed in natural language or in weakly typed languages like Python in LLMs' pretraining corpora. The LLM blends conventions of planning and Java test case generation, resulting in a syntactically wrong test case. By decomposing complex tasks and assigning each sub-task to a specialized LLM agent, we prevent this interference and improve the effectiveness of \method in unit test generation.

\paragraph{Mitigating Hallucination via a Panel Discussion}
LLMs are notoriously prone to hallucination, generating plausible yet nonfactual content~\cite{Huang2023AQuestions}. We found that hallucination is particularly frequent in test-oracle generations, where LLMs produce incorrect oracles even when clear instructions are provided. For instance, when told to \texttt{square \allowbreak the \allowbreak integer \allowbreak  if its index is a multiple of 3}, an LLM may incorrectly leave the number unchanged. To address this, we use multiple LLMs to independently generate oracles and then compare their outputs. Since hallucinations are usually inconsistent, taking a consensus across models helps filter them out. Instead of simple majority voting, we allow the LLMs to explain their reasoning in a panel discussion and determine the answer based on the most consistent logic.

\paragraph{Stopping Overthinking via a Dual-LLM Pipeline} Reasoning LLMs like DeepSeeek R1 are effective but often overly verbose, producing lengthy outputs that significantly increase generation time. In our experiment, using DeepSeek directly for oracle evaluation sometimes produced outputs exceeding \num{10000} tokens, taking hours to complete a single test file. In fact, we observed that DeepSeek often identified correct oracles early. Still, the model would continue second-guessing itself by thinking ``\texttt{I think the correct oracle should be 147. But, wait, maybe I made a mistake ...}''. To prevent this ``overthinking'', we truncate DeepSeek's output at \num{2000} tokens and use a basic LLM to extract the correct oracle from its reasoning. This dual-LLM pipeline preserves the benefit of deep reasoning while keeping outputs concise.

\subsection{Threats to validity}
\label{subsec:threats}
\noindent\textbf{Construct Validity.} We evaluate \method using code coverage, mutation score, and oracle correctness. While real bug detection on datasets like Defects4J is another useful metric, we did not include it because such project-level datasets are beyond the scope of this work, as discussed earlier. Instead, we use the mutation score as a proxy for bug-finding capability.

\noindent\textbf{Internal Validity.} Our results may be influenced by the choice of LLMs used in the framework. We use LLaMA 3.1 70B as the basic LLM and DeepSeek R1 as the reasoning LLM, selected based on preliminary experiments for their stability and open-source availability. While other popular models, such as GPT, Grok, and Mistral, exist, exhaustively evaluating all LLMs is beyond the scope of this work. The choice of LLMs may affect performance, but we aim to demonstrate the potential effectiveness of our method rather than benchmark specific models, which might indeed yield even better results. Another potential threat is data leakage, where our datasets may be included in the LLMs' pretraining data, leading to inflated performance. To mitigate this, we evaluate the mutation score, which detects behavioral changes in synthetically modified programs. These mutated versions are highly unlikely to appear in pretraining corpora, providing a more reliable evaluation of \method.

\noindent\textbf{Conclusion Validity.} \rev{A potential threat is the randomness in LLM outputs, which may affect the consistency of results. Specifically, LLMs can produce different outputs even when given identical inputs. To mitigate this, we repeat all experiments three times, compute the average, and apply statistical testing to assess the significance of differences in averages across programs.}

\noindent\textbf{External Validity.} One threat is that our evaluation is conducted on only two datasets (i.e., \human and LeetCodeJava). It is unclear how well \method and the baselines (\evo, \empirical, and \togll) generalize to other datasets, especially those that involve complex programs with dependencies on user-defined or external classes. However, \human is a representative, commonly used benchmark dataset in recent studies~\cite{jiang2023impact,Siddiq2024UsingStudy,chen2024chatunitest,yuan,kang2025explainable,li2024hybrid,endres2024can,tian2025fixing}, while we created LeetCodeJava to demonstrate the effectiveness of \method on programs with higher and varying complexity. Moreover, our experimental results show that automated test generation, especially oracle generation, remains unresolved by existing approaches on these datasets. Our work thus represents an important first step towards generating tests for more complex programs. In fact, \method is a multi-agent framework, which can be readily extended by introducing a dedicated agent to handle dependencies. Such an agent would retrieve definitions of user-defined or external classes, augmenting the test generation process when complex dependencies are involved.  We leave this exploration for future work. Last, the strong performance of \method on the LeetCodeJava dataset, which contains problems of varying difficulty, suggests that it is robust across diverse testing scenarios.

\section{Related Work}
\label{sec:related_works}
We report prior work on test prefix generation and specification-based test oracle generation in Sections~\ref {subsec:related_prefix} and~\ref {subsec:related_oracle}, respectively. 
\subsection{Test Prefix Generation}
\label{subsec:related_prefix}
Early works in what is commonly referred to as test case generation are, in fact, more accurately characterized as test prefix generation accompanied by regression oracles. These works in test prefix generation primarily aim to automate manual testing by generating test prefixes with high code coverage. Traditional techniques in this area include fuzzing~\cite{Miller1990AnUtilities,Fioraldi2023DissectingEvaluation}, feedback-directed random test generation ~\cite{Csallner2004JCrasher:Java,Pacheco2007Randoop:Java,Pacheco2008FindingTesting,Selakovic2018TestLanguages,Arteca2022Nessie:Callbacks}, dynamic symbolic execution~\cite{Godefroid2005DART:Testing,Sen2005CUTE:C,Cadar2008EXE:Death,Tillmann2014TransferringDigger}, and search/evolutionary algorithm-based approaches~\cite{Fraser2011EvolutionarySuites,Fraser2011EvoSuite:Software,Pacheco2007Randoop:Java}. Among these approaches, \evo has been extensively studied and applied to more than a hundred open-source software projects and several industrial systems, identifying thousands of potential bugs. However, as noted by ~\citet{zhang2025large}, traditional approaches such as \evo often struggle to produce meaningful, human-readable test cases, despite their remarkable code coverage. This is primarily due to their limited understanding of the semantics of the focal methods. To alleviate such issues, recent research has turned to LLMs to produce more practical test prefixes with high readability and coverage.

\citet{Tufano2021UnitContext} introduced the first LLM-based unit test generation approach AthenaTest, which formulates the task as a sequence-to-sequence learning problem.  AthenaTest first denoises the pre-training of LLMs on a large Java corpus and then performs supervised fine-tuning for the downstream task of test generation. This work inspired subsequent research on fine-tuning LLMs for test generation ~\cite{ALAGARSAMY2024107565,catlm,unitysyn}. Additionally, researchers have explored various strategies to improve LLM training for test case generation, including reinforcement learning~\cite{reinforce}, domain adaptation~\cite{domain}, and data augmentation~\cite{aug}.

Alternatively, a myriad of works perform prompt engineering using off-the-shelf LLMs, following a \emph{generation-and-refinement} paradigm, where initial test cases are first generated based on prompts and then iteratively refined using dynamic execution feedback (e.g., code coverage report and failure information) ~\cite{chen2024chatunitest,gu2024testart,ni2024casmodatest,altmayer2025coverup, codeaware,Schafer2024AnGeneration,wang2024hits,yuan,zhang2025large}. To further improve the quality of test prefixes, researchers propose a wide range of strategies to incorporate LLMs with valuable contextual information, including mutation testing~\cite{dakhel2024effective}, method slicing ~\cite{wang2024hits}, demonstration retrieval ~\cite{zhang2024llm}, defect detection ~\cite{yin2025you}, and program analysis ~\cite{yang2024enhancing}. 

Despite the rapid growth in LLM-based test generation research, ~\citet{tang2024chatgpt} reported that \evo remains the SOTA approach in generating test prefixes with high code coverage. \method builds on prompt engineering-based test generation and makes the following contributions: (1) \method is the first multi-agent LLM-based test generator. \method lays the foundation for multi-agent frameworks in test generation by proposing a generic, effective framework with carefully designed agents, including \emph{Planner}, \emph{Tester}, and \emph{Inspector}. (2) \method presents a breakthrough in LLM-based test prefix generation, reaching comparable code coverage and significantly better mutation rates than evo. 

\subsection{Test Oracle Generation}
\label{subsec:related_oracle}
Test oracle generation requires an accurate understanding of program semantics and requirements, posing significant challenges for early automated testing tools~\cite{chen2024chatunitest,gu2024testart,ni2024casmodatest,altmayer2025coverup, codeaware,Fraser2011EvoSuite:Software,Pacheco2007Randoop:Java}. These tools circumvent these challenges by generating regression oracles, while specification-based oracle generation has been underexplored. Regression oracles are derived from the current implementation and are effective for detecting behavioral changes across software versions; however, they are limited in validating functional correctness against intended specifications. This limitation arises because regression oracles assume the existing code is correct, potentially masking bugs rather than detecting them. 

However, going beyond regression oracles and generating specification-based oracles is challenging because it requires a deep understanding of functional requirements, edge cases, and expected outcomes~\cite{binta2024doc2oracle}. With the advent of LLMs, researchers have begun to explore their potential for generating accurate specification-based oracles, using both prompt engineering and fine-tuning strategies. \citet{Dinella2022Toga:Generation} introduces TOGA, a transformer-based approach for inferring specification-based oracles from the context of the focal method. TOGA is the first LLM-based test oracle generator powered by a fine-tuned CodeBERT LLM. Inspired by this work, several studies have proposed generating specification-based oracles using either FT or PE strategies ~\cite{chatassert,Hossain2024TOGLL:LLMs,chatassert,catlm,endres2024can,Siddiq2024UsingStudy}.  ~\citet{Siddiq2024UsingStudy} proposed \empirical, a solution using prompt engineering to generate JUnit tests based on natural language descriptions, without fine-tuning. It focuses on generating both test prefixes and oracles using only prompt engineering. ~\citet{zhang2025exploring} conduct the first comprehensive study on fine-tuning various LLMs for oracle generation across two benchmarks, five LLMs, and two metrics. Building on this study, ~\citet{zhang2025improving} introduced RetriGen, a retrieval-augmented oracle generation approach that incorporates a novel hybrid assertion retriever into the oracle generation process. RetriGen retrieves most relevant assertions from external codebases by leveraging both lexical and semantic similarity.  TOGLL\cite{Hossain2024TOGLL:LLMs} combines \evo-generated test prefixes with LLM-generated oracles through fine-tuned models, achieving SOTA performance in oracle generation for Java programs. However, TOGLL depends on test prefixes generated by \evo and requires the extensive fine-tuning of large models, which can be resource-intensive and less adaptable to new domains.

\method also relies on prompt engineering-based oracle generation but introduces several key innovations, including the design of multiple specialized agents, the panel discussion to reduce hallucination, and the dual-LLM pipeline to extract structured information from the verbose output of reasoning LLMs.


\section{Conclusion}
\label{sec:conclusion}
In this work, we presented \method, a novel multi-agent, end-to-end framework for generating high-quality unit tests with accurate oracles. \method does not require fine-tuning or the use of external tools but leverages a panel discussion to mitigate hallucinations and employs a dual-LLM pipeline to reduce overthinking during reasoning.

Extensive evaluation on two benchmarks, HumanEvalJava and LeetCode, demonstrates that \method achieves performance comparable to \evo in terms of line coverage and branch coverage, while substantially outperforming \evo in mutation score. Additionally, \method produces accurate specification-based oracles that outperform the state-of-the-art oracle generator \togll by a large margin ($>=21.1$ percentage points). Ablation studies highlight the critical roles of key agents: the \emph{Planner} for enhancing test prefix quality, and the \emph{Requirement Engineer} together with the panel discussion for improving oracle correctness. For future work, we plan to extend \method to project-level test generation and explore more advanced panel discussion strategies to further reduce hallucinations.
\section{Acknowledgement}
This work was supported, in part, by Science Foundation Ireland Grant 16/RC/3918 and by Huawei Technologies Co., Ltd. Lionel Briand is supported by the Natural Sciences and Engineering Research Council of Canada.
\bibliographystyle{ACM-Reference-Format}
\bibliography{references}

\newpage
\section*{Appendix}
\subsection*{\method with Alternative LLMs}
\begin{table}[hbt]
    \centering
    
    \begin{tabular}
    {
        @{\hspace{0.5\tabcolsep}}c
        @{\hspace{0.5\tabcolsep}}c
        @{\hspace{0.5\tabcolsep}}c
        @{\hspace{0.5\tabcolsep}}c
        @{\hspace{0.5\tabcolsep}}c
    }
    \toprule
    Dataset & Metric & Llama & CodeLlama & Mistral \\
    \midrule
    \multirow{5}{*}{HumanEvalJava} & Line & \textbf{0.991} & 0.991 & 0.840* \\
    & Branch & \textbf{0.970} & 0.967 & 0.839* \\
    & Mutation & \textbf{0.980} (2384/2443) & 0.980 (2390/2443) & 0.796* (1945/2443) \\
    & Oracle & 0.910 & \textbf{0.912} & 0.861* \\
    \midrule
    \multirow{5}{*}{Leetcode-Medium} & Line & \textbf{0.990} & 0.990 & 0.914* \\
    & Branch & 0.949 & \textbf{0.950} & 0.943* \\
    & Mutation & 0.939 (574/611) & \textbf{0.948} (579/611) & 0.839* (513/611) \\
    & Oracle & \textbf{0.930} & 0.919 & 0.904* \\
    \midrule
    \multirow{5}{*}{Leetcode-Hard} & Line & 0.989 & \textbf{0.990} & 0.876* \\
    & Branch & 0.980 & \textbf{0.981} & 0.871* \\
    & Mutation & \textbf{0.937} (805/859) & 0.930 (799/859) & 0.905* (777/859) \\
    & Oracle & \textbf{0.894} & 0.869* & 0.848* \\
    \bottomrule

    \end{tabular}
    \caption{Experimental results of \method using alternative LLMs. ``Llama'', ``CodeLlama'', and ``Mistral'' denote \method using \textit{Llama 70B},\textit{ CodeLlama 70B}, and \textit{Mistral 22b}, respectively. The highest value in each row is highlighted in bold. The ``*'' indicates that the difference between this result and \method using Llama 70B is statistically significant.}
    \label{tab:alterllm}
\end{table}

\rev{Table 3 reports the experimental results of \method using alternative LLMs. We observe that Mistral 22B has the lowest effectiveness among the three models, resulting in substantially lower line coverage, branch coverage, mutation scores, and oracle correctness. Wilcoxon Signed Rank tests confirm that all differences between Mistral 22B and Llama 70B are statistically significant, with p-values below $10^{-4}$. This outcome is expected, as Mistral 22B has considerably fewer parameters than Llama 70B and CodeLlama 70B, making it less capable across tasks, including test generation.}

\rev{In contrast, CodeLlama 70B and Llama 70B yield comparable results. For test prefix generation, the largest difference between the two models is a 0.09 gap in mutation score on the LeetCode-Medium dataset. However, Wilcoxon Signed Rank tests show that none of these differences are statistically significant. For oracle generation, Llama 70B achieves slightly higher correctness on LeetCode-Medium ($0.011 = 0.930 - 0.919$) and LeetCode-Hard ($0.025 = 0.894 - 0.869$), while performing slightly worse on HumanEvalJava ($0.002 = 0.912 - 0.910$). The differences between HumanEvalJava and LeetCode-Medium are not statistically significant, whereas the difference between LeetCode-Hard and HumanEvalJava is. Overall, Llama 70B performs marginally better than CodeLlama 70B. For this reason, we select Llama 70B as the default model for \method and report the corresponding results in the main manuscript.}

\end{document}